\documentclass[aps,twocolumn,amsmath,amssymb,showpacs,prl]{revtex4}
\usepackage{epsf}
\usepackage{graphicx}
\usepackage{gensymb}
\usepackage{sidecap}

\newcommand{\etal}{{\it et al.}}

\begin{document}

\title{Importance~of~Fermi~surface~topology~for~high~temperature~superconductivity
in~electron-doped~iron~arsenic~superconductors}

\author{Chang~Liu}
\affiliation{Ames Laboratory and Department of Physics and
Astronomy, Iowa State University, Ames, Iowa 50011, USA}
\author{A.~D.~Palczewski}
\affiliation{Ames Laboratory and Department of Physics and
Astronomy, Iowa State University, Ames, Iowa 50011, USA}
\author{Takeshi~Kondo}
\affiliation{Ames Laboratory and Department of Physics and
Astronomy, Iowa State University, Ames, Iowa 50011, USA}
\author{R.~M.~Fernandes}
\affiliation{Ames Laboratory and Department of Physics and
Astronomy, Iowa State University, Ames, Iowa 50011, USA}
\author{E.~D.~Mun}
\affiliation{Ames Laboratory and Department of Physics and
Astronomy, Iowa State University, Ames, Iowa 50011, USA}
\author{H.~Hodovanets}
\affiliation{Ames Laboratory and Department of Physics and
Astronomy, Iowa State University, Ames, Iowa 50011, USA}
\author{A.~N.~Thaler}
\affiliation{Ames Laboratory and Department of Physics and
Astronomy, Iowa State University, Ames, Iowa 50011, USA}
\author{J.~Schmalian}
\affiliation{Ames Laboratory and Department of Physics and
Astronomy, Iowa State University, Ames, Iowa 50011, USA}
\author{S.~L.~Bud'ko}
\affiliation{Ames Laboratory and Department of Physics and
Astronomy, Iowa State University, Ames, Iowa 50011, USA}
\author{P.~C.~Canfield}
\affiliation{Ames Laboratory and Department of Physics and
Astronomy, Iowa State University, Ames, Iowa 50011, USA}
\author{A.~Kaminski}
\affiliation{Ames Laboratory and Department of Physics and
Astronomy, Iowa State University, Ames, Iowa 50011, USA}

\date{\today}
\begin{abstract}
We used angle resolved photoemission spectroscopy and thermoelectric
power to study the poorly explored, highly overdoped side of the
phase diagram of Ba(Fe$_{1-x}$Co$_x$)$_2$As$_2$ high temperature
superconductor. Our data demonstrate that several Lifshitz
transitions - topological changes of the Fermi surface - occur for
large $x$. $T_c$ starts to decrease with doping when the
cylindrical, central hole pocket changes to ellipsoids centering at
the $Z$ point, and goes to zero before these ellipsoids disappear
around $x=0.2$. Changes in thermoelectric power occur at similar
$x$-values. Beyond this doping level the central pocket changes to
electron-like and superconductivity does not exist. Our observations
reveal the crucial importance of the underlying Fermiology in this
class of materials. A necessary condition for superconductivity is
the presence of the central hole pockets rather than perfect nesting
between central and corner pockets.
\end{abstract}


\maketitle

The phase diagrams of the newly-discovered iron arsenic
superconductors contain a number of intriguing features. For the
electron-doped $A$(Fe$_{1-x}T_x$)$_2$As$_2$ series (122, $A$ = Ca,
Sr, Ba; $T$ = Co, Ni, Pd, etc.), superconductivity is found in both
regions with and without a long-range antiferromagnetic (AFM) order
\cite{NiNiCo, Fisher, Ning, Fang, Nandi, Canfield, Canfield_Review}.
The superconducting (SC) region extends to different doping levels
for different dopants, but scales very well if the horizontal axis
of the phase diagram was chosen to be the number of extra electrons
\cite{Canfield, Canfield_Review}. It is therefore likely that
changes in the underlying electronic structure due to electron
doping are linked closely to their SC behavior. On the underdoped
side, a recent angle-resolved photoemission spectroscopy (ARPES)
study on Ba(Fe$_{1-x}$Co$_x$)$_2$As$_2$ \cite{ChangNPhys} revealed
that superconductivity emerges at a doping level ($x_{\textrm{on}}$)
where a topological change of the Fermi surface (Lifshitz transition
\cite{Lifshitz} at doping $x_1$) reduces the magnetically
reconstructed Fermi surface to its paramagnetic appearance, i.e.
$x_1 \simeq x_{\textrm{on}}$. This transition exhibits itself as a
rapid change of Hall coefficient and thermoelectric power (TEP) in
transport measurements \cite{Mun}. An immediate question is whether
a similar change of Fermiology causes the collapse of the SC dome on
the heavily overdoped regime. It is inevitable that the hole pockets
surrounding the central axis of the Brillouin zone ($\Gamma$-$Z$)
will shrink in size and vanish at some higher doping $x_2$. The
question is whether this Lifshitz transition correlates with the
offset of superconductivity on the overdoped side of the phase
diagram ($x_{\textrm{off}}$). Based on a solution of the two-band
BCS gap equations, assuming only interband coupling, Fernandes and
Schmalian \cite{Reafel_new} showed that the disappearance of
superconductivity is directly linked to the vanishing of the central
hole pocket(s), i.e. $x_2\simeq x_{\textrm{off}}$. Experimentally,
the Hall coefficient vs. doping on Ba(Fe$_{1-x}$Co$_x$)$_2$As$_2$
\cite{Fang} experiences a slight change of slope around
$x_{\textrm{off}}$, hinting at a possible Lifshitz transition close
to the high doping offset of superconductivity.

\begin{figure}[b]
\includegraphics[width=3.5in]{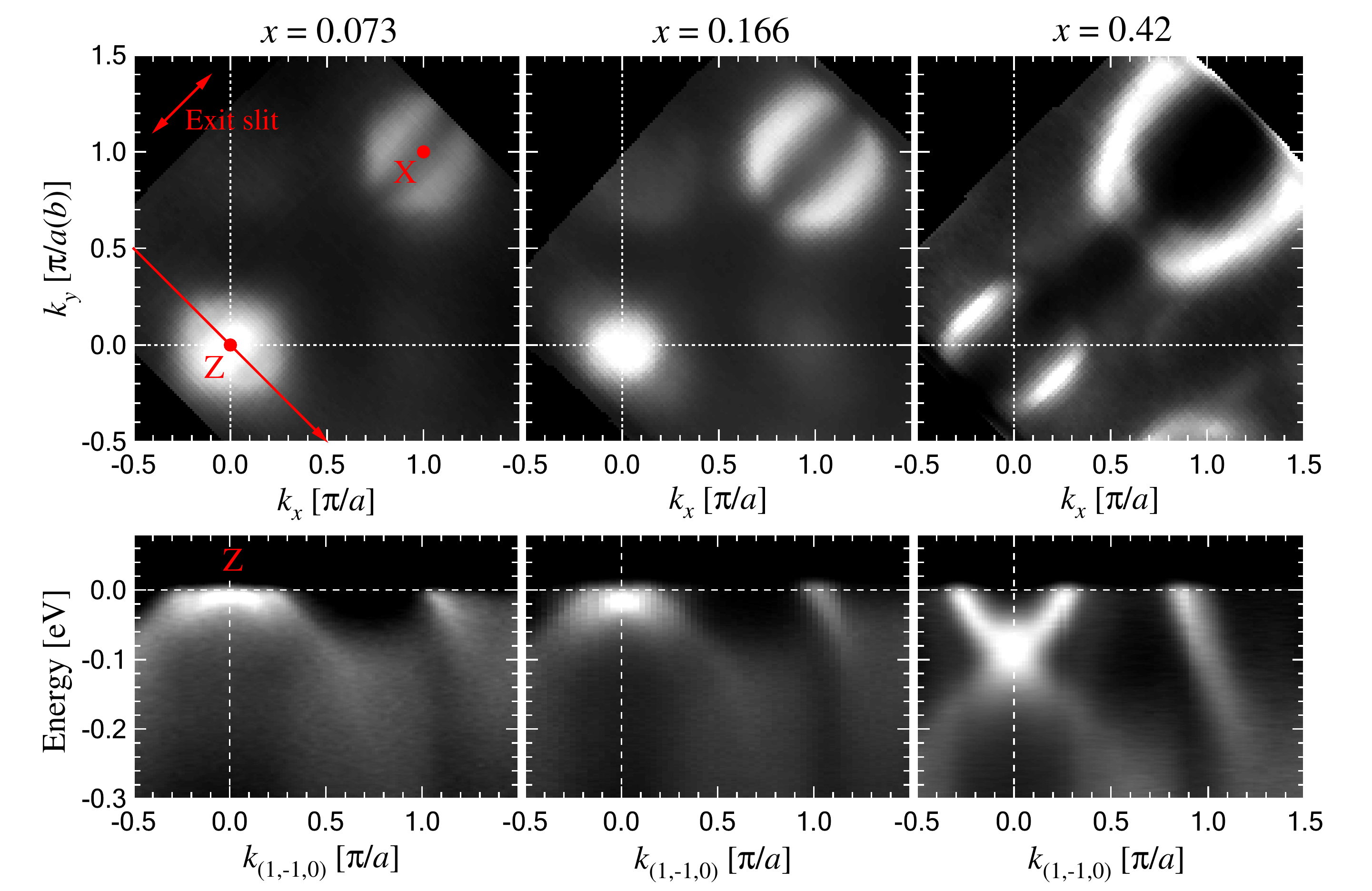}
\caption{(Color online) Fermi maps and band dispersion around the
upper zone edge $Z$ of Ba(Fe$_{1-x}$Co$_x$)$_2$As$_2$ for $x=0.073$
(optimal doping), $x=0.166$ (edge of SC dome) and $x=0.42$. Upper
Row: Fermi mappings for the three doping levels, taken with incident
photon energy $h\nu=35$ eV at temperature $T=20$ K. Red arrows show
the exit slit direction of the hemispheric analyzer and the cutting
direction of the band dispersion maps (Lower Row). The same
direction is also used in Fig. 2.}
\end{figure}

\begin{figure*}[t]
\includegraphics[width=7in]{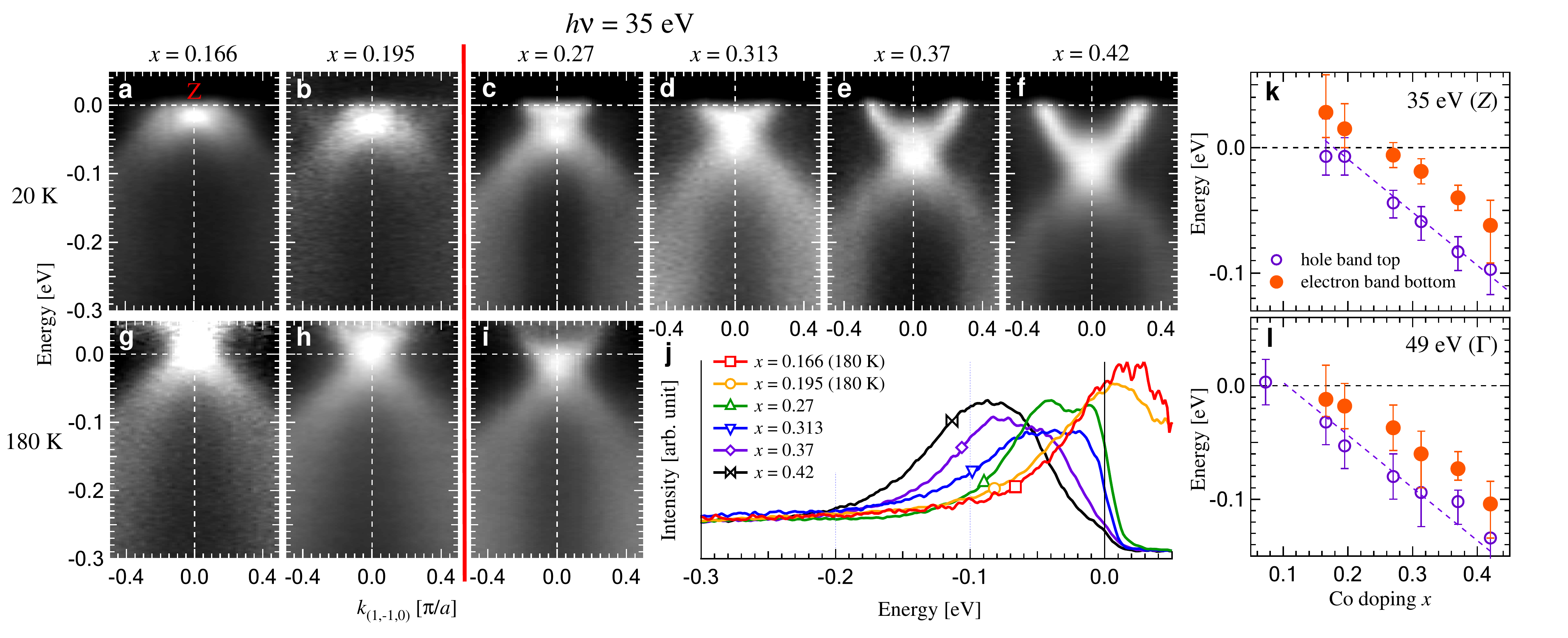}
\caption{(Color online) Band location analysis for the Lifshitz
transitions. (a)-(i): Band dispersion maps along the direction shown
in Fig. 1 for six different doping levels (top of each column) at
low and high temperatures (left of each row). All data is taken with
35 eV photons. The high temperature data is divided by the
resolution convoluted Fermi-Dirac function for better pinpointing
the band positions above the Fermi level. The red vertical line
indicates that a Lifshitz transition happens between $x = 0.195$ and
$x = 0.27$ at $Z$. (j) Energy distribution curves (EDCs) at $Z$ for
the doping levels in panels (a)-(i). Measurement temperature is
$T=20$ K unless specifically mentioned in the graph. (k),(l):
Evolution of binding energy for the top of the hole band and the
bottom of the electron band with respect to cobalt doping. Data is
extracted from ARPES intensity maps taken with (k) 35 eV and (l) 49
eV photons, corresponding to $k_z$ values of $Z$ and $\Gamma$
points, respectively. Data points in (k) are extracted from the EDCs
at (j) by fitting with two Lorentzian functions (times the Fermi
function for low temperature data). Raw data for extracting panel
(l) is not shown.}
\end{figure*}

In this Letter we study this issue in detail using ARPES and TEP
measurements. We performed a complete survey of the electronic
structure on the overdoped part of the phase diagram of this
material. This survey reveals that topological changes of the Fermi
surface link directly to superconductivity in electron-doped
pnictides. In the overdoped side, superconducting transition
temperature $T_c$ starts to be suppressed around the doping level
($x_{2\Gamma}$) where the cylindrical hole pocket surrounding the
zone center ($\Gamma$-$Z$) changes to ellipsoids centering at $Z$.
$T_c$ is driven to zero before the disappearance of $Z$-ellipsoids
and the change in TEP at $x_{2Z}\sim0.2$. In short, we find that
$x_{2\Gamma} < x_{\textrm{off}} < x_{2Z}$. Our data demonstrated
that superconductivity in the pnictides is very robust with respect
to doping; the whole $\Gamma$ Fermi sheet has to be almost
completely eliminated in order to drive $T_c$ to zero. A necessary
condition for superconductivity then is the existence of the central
hole pockets rather than a perfect nesting between the $\Gamma$ and
$X$ pockets \cite{Terashima}. The dominant contribution to the
pairing interaction is believed to come from inter-band coupling
\cite{Reafel_new}.

Single crystals of Ba(Fe$_{1-x}$Co$_x$)$_2$As$_2$ were grown out of
a self-flux using conventional high-temperature solution growth
techniques \cite{NiNiCo}. The doping level $x$ was determined using
wavelength dispersive X-ray spectroscopy in a JEOL JXA-8200 electron
microprobe \cite{NiNiCo}. Long range antiferromagnetism was observed
below a transition temperature $T_\mathrm{N}(x)$ up to $x \sim
0.06$. Superconductivity appears around $x_{\textrm{on}} = 0.038$
and vanishes between $0.135 < x_{\textrm{off}} \leq 0.166$ (see Fig.
4) \cite{Canfield_Review}. The ARPES measurements were performed at
beamline 10.0.1 of the Advanced Light Source (ALS), Berkeley,
California using a Scienta R4000 electron analyzer. Vacuum
conditions were better than $3\times10^{-11}$ torr. The energy
resolution was set at $\sim25$ meV. All samples were cleaved
\textit{in situ} yielding mirror-like, fresh $a$-$b$ surfaces. High
symmetry points were defined the same way as in Ref.
\cite{ChangNPhys}. TEP measurements were made as described in Ref.
\cite{Mun}.

Fig. 1 shows the ARPES Fermi maps and corresponding band dispersion
data for three different doping levels of
Ba(Fe$_{1-x}$Co$_x$)$_2$As$_2$ \cite{direction}. The incident photon
energy is $h\nu=35$ eV, corresponding to $k_z \simeq 2\pi/c$, the
upper edge of the first Brillouin zone ($Z$) \cite{Takeshi_nesting}.
From data in Fig. 1 it is clear that, as electron doping initially
increases, the Fermi contours around $Z$ shrink in size. At $x =
0.166$, the edge of the SC dome, the $Z$-pocket shrinks to almost a
single point, meaning a complete vanishing of the hole pocket. This
observation is consistent with the data in Refs. \cite{ChangNPhys,
Sekiba, Brouet}. As $x$ increases, the $Z$ pocket expends again,
yielding a diamond shape at $x = 0.42$. Band dispersion clearly
reveals that this ``diamond" is electron-like. Such an electron
pocket is not predicted by band structure calculations \cite{Fang}.
The $X$ pocket, on the other hand, keeps expanding from $x = 0.073$
to $x = 0.42$, and it remains electron-like. The central message of
this figure is that the $Z$-pocket undergoes a drastic topological
change from hole-like to electron-like at roughly the doping level
where superconductivity vanishes. Based on this observation we
perform two independent data analysis procedures with finer doping
steps to further pinpoint the doping level at which the Lifshitz
transition takes place.

\begin{figure}
\centering
\includegraphics[width=3.5in]{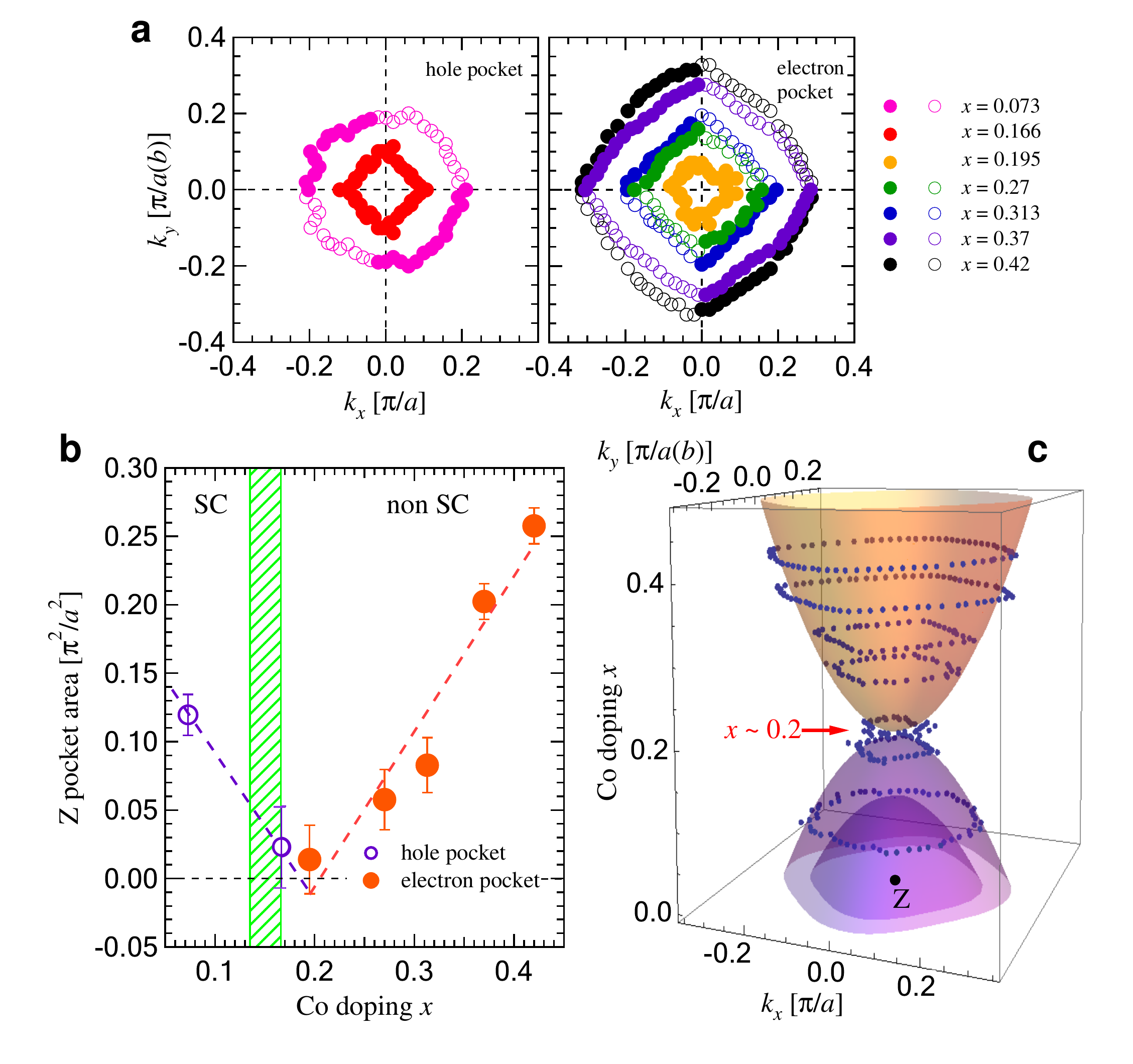}
\caption{(Color online) Pocket size analysis for the Lifshitz
transition at upper zone boundary $Z$. (a) $Z$ pocket extraction for
seven doping levels, done by fitting the momentum distribution
curves (MDCs) at the chemical potential with several Lorenzian
functions. Positions of hollow circles are symmetrized from
experimental data points (solid circles), proposing the band
positions where ARPES intensity is suppressed by the transition
matrix element. (b) Evolution of $Z$ pocket area with cobalt doping.
Green shaded area indicates the boundary of the SC dome. (c)
Visualization of the Lifshitz transition. Data in (a) is plotted
against the cobalt doping $x$ as a third dimension. Shaded areas are
approximate size and shape of the pockets. Panels (b) and (c) show a
Lifshitz transition at $x_{2Z}\sim0.2$.}
\end{figure}

First, to obtain a more accurate value for $x_2$, we extract the
energies for the hole band top and the electron band bottom at the
zone center, and examine them as a function of cobalt doping. As
shown in Fig. 2, we plot the band dispersion maps along the same
direction as in Fig. 1 for six different doping levels ranging from
$x = 0.166$ to $x = 0.42$, and use the energy distribution curves
(EDCs) in Fig. 2(j) \cite{Temp} to see that both the hole band and
the electron band shift to higher binding energies as $x$ increases.
The shape of these bands remain the same during the process. There
is a small gap ($\sim25$ meV) between these two bands. At $0.195 < x
< 0.27$ the bottom of the electron band moves above the Fermi level,
as revealed in Fig. 2(k) where two Lorentzians are fitted to the
data from Fig. 2(j). At a slightly lower doping level the top of the
hole pocket also moves above $E_F$. Figs. 2(a)-(k) illustrate that,
at the $Z$ point of the Brillouin zone, the Lifshitz transition
takes place between $0.195 < x_{2Z} < 0.27$, higher than
$x_{\textrm{off}} \sim 0.15$.

The intrinsic three dimensionality of the electronic structure
\cite{ChangLiu_3D, Vilmercati} results in different $x_2$ values for
different $k_z$. In Fig. 2(l) we investigate this effect by
performing the same analysis to the data taken with 49 eV photons
(raw data not shown). This incident photon energy corresponds to
$k_z \simeq 0$, the central point of the Brillouin zone ($\Gamma$).
We see that indeed the Lifshitz transition shifts to a lower doping,
i.e. $x_{2\Gamma} \sim 0.11$. We note that this is the doping level
where $T_c$ starts to decrease in the phase diagram. This
observation also supports the theoretical prediction that three
dimensionality of the Fermi surface leads to a more gradual decrease
of $T_c$ in the overdoped side \cite{Reafel_new}.

In Fig. 3 we perform a pocket size analysis at $Z$ to further
pinpoint $x_2$. This second procedure is independent from the above
energy extraction method. To do this we first find the $Z$ pocket
location for seven doping levels (ranging from $x = 0.073$ to $x =
0.42$) by fitting the momentum distribution curves (MDCs) at the
chemical potential with several Lorenzian functions. From Fig. 3(a)
we see a clear evolution of the $Z$ pocket size with doping. As $x$
increases, the hole pocket shrinks in size up to $x = 0.195$. Above
this doping an electron pocket appears and increases in size up to
the highest doping measured. As seen in Fig. 3(b), both the hole and
electron pocket size evolves in a linear fashion, a signature of the
validness of the rigid band shifting scenario \cite{Neupane}, and of
the pockets being paraboloids in shape. The cross-over takes place
around $x = 0.2$. This Lifshitz transition is best visualized in
Fig. 3(c) where data in Fig. 3(a) is plotted against the cobalt
doping $x$ as a third dimension. This figure reveals that, as cobalt
concentration increases, the Fermi sea level rises and the $Z$ hole
bands gradually drop below it. At $x \sim 0.2$ the total occupation
of the outer hole band marks the Lifshitz transition. Beyond this
point the $Z$ pocket changes to electron-like, and superconductivity
vanishes.

\begin{figure}[t!]
\includegraphics[width=3.4in]{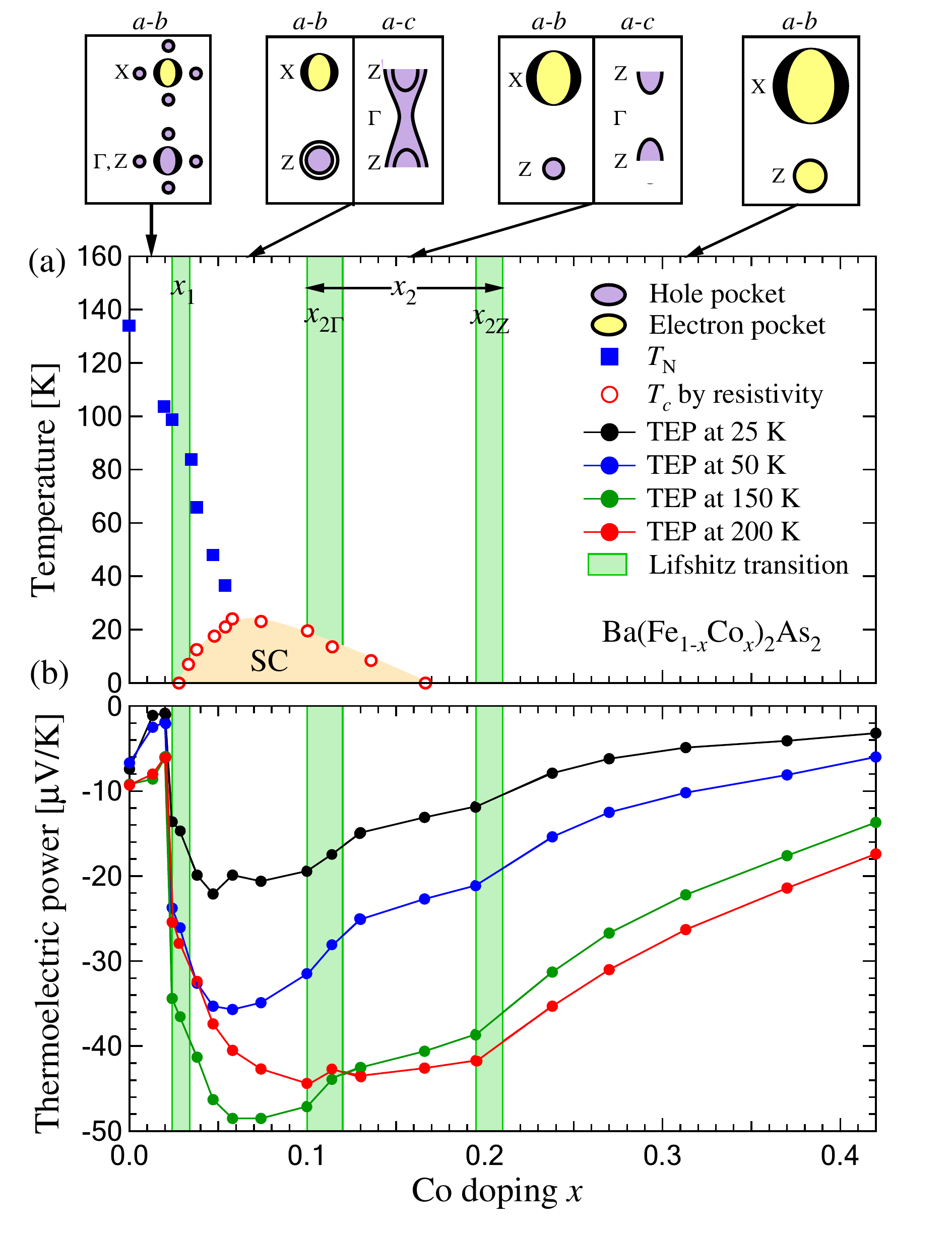}
\caption{(Color online) (a) location of the known Lifshitz
transitions in the phase diagram. $T_\mathrm{N}$ and $T_c$ data is
taken from Refs. \cite{NiNiCo} and \cite{ChangNPhys}. Top insets
show schematic Fermi surface topology in the $a$-$b$ and $a$-$c$
plane for each region in the phase diagram. (b) Thermoelectric power
vs. doping for four different temperatures.}
\end{figure}

Fig. 4 summarizes our systematic ARPES survey on the Fermi surface
topology of Ba(Fe$_{1-x}$Co$_x$)$_2$As$_2$ for $0 \leq x \leq 0.42$
and compares it with TEP data over the same doping range. The most
important finding of this study is that the low- and high-doping
onset of the SC region link closely to topological changes of the
Fermi surface. The first Lifshitz transition at the low doping onset
of superconductivity is described in detail in Ref.
\cite{ChangNPhys, Mun}. The second and third Lifshitz transitions
occur for $0.11 \lesssim x \lesssim 0.2$ and correspond to the high
doping onset of superconductivity. $x_{2\Gamma} \simeq 0.11$
corresponds to the doping level where the shape of the
quasi-cylindrical outer $\Gamma$ contour changes to an ellipsoid
centering at $Z$. Also at $x \simeq 0.11$ the superconducting $T_c$
starts to suppress. As doping is increased, this $Z$ ellipsoid
shrinks in size until it disappear altogether at $x_{2Z} \simeq
0.2$. On the other hand, superconductivity vanishes at
$x_{\textrm{off}} \simeq 0.15$. At $x
> 0.2$, the region of the highest doping, the central pocket changes
to electron-like, and superconductivity does not exist. Our TEP
data, plotted as $S(x)|_{T = \textrm{const}}$ for several
temperatures in Fig. 4(b), show clear step-like or change-of-slope
anomalies at Co-concentrations that are in an excellent agreement
with those at which the the Lifshitz transitions were detected by
ARPES [Fig. 4(a)]. These results, taken together, confirm extreme
sensitivity of TEP to the changes in FS topology \cite{Blanter}.

Importantly, the above conclusion most likely also applies to other
electron doped 122 systems. We are specially interested in
\textit{A}(Fe$_{1-x}$Ni$_x$)$_2$As$_2$ where each nickel atom gives
two extra electrons per Fe site compared to one in the cobalt doped
system \cite{Canfield_Review}. There, similar to the cobalt doped
system, the Hall coefficient and thermoelectric power anomaly occurs
right at the onset of superconductivity \cite{ChangNPhys, Butch}.
Based on a similar ARPES survey \cite{Ari} we indeed find Lifshitz
transitions at close vicinity to the boundaries of
superconductivity, the only difference being that the corresponding
doping levels are roughly one half as those of the cobalt system. As
the phase diagram changes to $T$ vs. $e$, the extra electron count,
these two systems match perfectly.

Our findings have important implications on the nature of
superconductivity of the pnictides. First, our observation reveals
the crucial importance of the underlying Fermi surface topology: a
necessary condition for the emergence of superconductivity is the
existence of the non-reconstructed central hole pockets rather than
a perfect nesting condition between the central and corner pockets.
Superconductivity is not supported only when either one set of these
pockets (central or corner) vanishes, changes its carrier nature or
shows considerable reconstruction. Second, our results imply that
the suppression of superconductivity on the underdoped side is
related to the competition between the AFM and SC phases
\cite{Canfield_Review}, whereas on the overdoped side the
disappearance of the central hole pocket plays a more important role
than the decrease of the pairing interaction magnitude
\cite{Reafel_new}. Electron doped 122 systems are, therefore, clear
examples of high temperature superconductors whose superconducting
behavior is controlled primarily by the underlying Fermiology.

We thank Sung-Kwan Mo and Makoto Hashimoto for their grateful
instrumental support at the ALS. Ames Laboratory was supported by
the Department of Energy - Basic Energy Sciences under Contract No.
DE-AC02-07CH11358. ALS is operated by the US DOE under Contract No.
DE-AC03-76SF00098.


\begin{thebibliography}{99}
\bibitem{direction}
Direction of the band dispersion maps is perpendicular to the
exit-slit because the ARPES intensity of the $Z$ electron pocket
along the exit-slit direction is suppressed by the transition matrix
element.
\bibitem{Temp}
In Fig. 2(j) high temperature data divided by the Fermi function is
used for $x=0.166$ and $x=0.195$ in order to better locate the band
positions above $E_F$.
\end{thebibliography}
\end{document}